\documentstyle[12pt]{article}
\newcommand{\beq}{\begin{equation}}
\newcommand{\ber}{\begin{eqnarray}}
\newcommand{\eeq}{\end{equation}}
\newcommand{\eer}{\end{eqnarray}}
\begin{document}
\rightline{[SUNY BING 9/29/96]}
\vspace{1mm}
\begin{center}
{\LARGE \bf TESTING $CP$, $T$ AND $(V-A)$ SYMMETRY
THROUGH TAU LEPTONS}\\
\end{center}
\vspace{10mm}
\noindent
{Charles A. Nelson\footnote{Electronic address:
cnelson@bingvmb.cc.binghamton.edu \newline Invited talk at
WHEPP4 workshop, S.N. Bose Nat'l Centre for Basic Sciences,
Calcutta.}, Department of Physics, State
University of New York at Binghamton, Binghamton, N.Y.
13902-6016}\\
\vspace{10mm}
\begin{abstract}
Model independent tests for symmetry violations in tau decays
are important for determining whether the tau lepton is
elementary or, instead, macroscopic.  Such tests are also
significant steps towards resolving the outstanding $e - \mu
- \tau$ puzzle.  This talk reviews such tests in the context
of a general treatment of two-body $\tau$ decays which only
assumes Lorentz invariance and the treelike structure of the
$\tau^- \tau^+$ production-decay sequence.  Direct
measurement of polarized-partial-widths and of associated
``longitudinal-transverse W-exchange" interference
intensities will provide significant probes for
distinguishing elementary/macroscopic theories of the Higgs
and of the  \newline $\tau$, $\nu_{\tau}$.  The analogous
goals and
techniques apply to top quark decays.
\end{abstract}
\vskip 40pt

\section{INTRODUCTION}

The most outstanding open questions concerning the tau lepton
are \newline (i) Does the $\tau$ differ in some way from
simply being
a more massive version of the $\mu$ and $e$ ?  \newline (ii)
{\bf The $e - \mu - \tau$ puzzle:}  What
is the fundamental relationship between the $\tau$, $\mu$ and
$e$ ?  \newline Naively, differences among the charged
leptons might
be expected to be most easily observed for the  $\tau$ since
it is the most massive. One purpose of this talk is to
discuss a general treatment of two-body tau decays [1] which
only assumes Lorentz
invariance and exploits the tree-like structure of the
dominant contributions to the $\tau ^{-}\tau ^{+}$
production-decay sequence. In particular, $CP$ invariance,
$T$ invariance, and a $(V - A) $ structure of the
tau charged-current is not assumed.  At present, these
symmetries are
poorly tested for the $\tau$.  Discovery of a violation of
one of these symmetries in reactions involving the $\tau$,
might point to a more fundamental structure underlying the
charged leptons, e.g. be evidence for lepton compositeness.
For example, in analogy with the Pauli anomalous magnetic
moment, such structure could show up as an additional
tensorial $g_{+}=f_M+f_E$ coupling
which would preserve the 3 signatures for only $\nu _L$
couplings but give non-($V-A$)-values to the semi-leptonic
parameters. We find that by the $\rho
^{-}$ ($a_1^{-}$) modes, compositeness in
the tau lepton
could be respectively probed [2] to $1.2TeV$($1.5TeV$).
In this talk we do not discuss the important tests for
symmetry violations in the neutral current couplings
involving the $\tau$, see [3].

The analogous goals and techniques apply to top quark decays
[4].

\section{A GENERAL TREATMENT OF $\tau$ SEMILEPTONIC DECAYS BY
POLARIZED-PARTIAL-WIDTH \newline MEASUREMENTS}

The goals of a general parametrizatin of two-body $\tau$
decays are (a) to determine the ``complete Lorentz structure"
of ${J^{Charged}}_{Lepton}$ directly from experiment, and (b)
to test in a model independent manner for the presence of
``additional Lorentz couplings".  Simultaneously, there are
simple tests for leptonic $CP$ violation and for leptonic $T$
violation in $\tau$ decays.

The physical idea is very simple:  We introduce 8 parameters
to describe the most
general spin-correlation function for the decay sequence $
Z^o,\gamma
^{*}\rightarrow \tau ^{-}\tau ^{+}\rightarrow (\rho ^{-}\nu
)(\rho ^{+}\bar
\nu )$\ followed by $\rho ^{ch}\rightarrow \pi
^{ch}\pi ^o$ including both
$\nu _L$,\ $\nu _R\ $helicities and both   $\bar \nu _R$,\
$\bar \nu _L\
$helicities.  Thus, by including the $\rho $ polarimetry
information that is available from the $%
\rho^{ch}\rightarrow \pi^{ch}\pi ^o$ decay distribution, the
polarized-partial-widths for $\tau ^{-}\rightarrow
\rho ^{-}\nu \ $ are directly measureable.  For instance,
the general angular distribution for
polarized $\tau^{-}_{L,R} \rightarrow \rho ^{-}\nu
\rightarrow
(\pi ^{-}\pi
^o)\nu $ is described by
\beq
\frac{dN}{d(\cos \theta_1^\tau )d(\cos \tilde \theta _a)d
\tilde \phi_a} =
{\bf n}_a[1\pm {\bf f}_a\cos \theta _1^\tau ]\mp (1/\sqrt{%
2})\sin \theta _1^\tau \sin 2\tilde \theta _a\ {\cal R}_\rho
[\omega \cos
\tilde \phi
_a+\eta ^{\prime }\sin \tilde \phi _a]
\eeq
with upper(lower) signs for a L-handed $\tau^-$ (R-handed),
where
\ber
\begin{array}{c}
{\bf n}_a=\frac 1{8}(3+\cos 2\tilde \theta _a+\sigma {\cal
S}_\rho [1+3\cos
2\tilde
\theta _a]) \\ {\bf n}_a{\bf f}_a=\frac 1{8}(\xi [1+3\cos
2\tilde
\theta _a]+\zeta {\cal S}_\rho [3+\cos 2\tilde \theta _a])
\end{array}
\eer
In this expression, $\cos \theta_1^\tau $ describes the
direction of the $\rho^-$ momentum in the $\tau^-$ rest
frame, and $\cos \tilde \theta _a$, and $\tilde \phi_a$
describe the direction of the $\pi^-$ in the $\rho^-$ rest
frame. Such formulas for more general spin-correlation
functions in terms of the 8 semi-leptonic parameters are
given in [1] for unpolarized $e^- e^+$ beams, and in [5] for
polarized beams.

There are eight $\tau$ semi-leptonic decay parameters since
there are the four $\rho_{L,T} \nu_{L,R}$ final states:
The first parameter is simply $\Gamma \equiv
\Gamma
_L^{+}+\Gamma _T^{+}$, i.e. the (full) partial width
for $\tau ^{-}\rightarrow \rho ^{-}\nu $.  The second is the
chirality
parameter $\xi \equiv \frac 1\Gamma (\Gamma _L^{-}+\Gamma
_T^{-})$.
Equivalently, \newline $ \xi \equiv$ (Prob $\nu_{\tau}$ is
$\nu_L$) $ - $
(Prob $\nu_{\tau}$ is $\nu_R$), or
\begin{equation}
\xi \equiv |< \nu_L |\nu_{\tau} >|^{2} - |< \nu_R |\nu_{\tau}
>|^{2}
\end{equation}
So a value $\xi = 1$ means the coupled $\nu_{\tau}$ is
pure $\nu_L$.  $\nu_L$ ($\nu_R$) means the emitted neutrino
has
L-handed (R-handed) polarization.
For the special case of a mixture of only $V$ \& $A$
couplings and $m_{
\nu_{\tau} } = 0 $, $\xi \rightarrow \frac{\left| g_L\right|
^2-\left|
g_R\right| ^2}{\left| g_L\right| ^2+\left|g_R\right| ^2}$ and
the ``stage-one spin correlation" parameter $\zeta
\rightarrow \xi$.
The subscripts on the $\Gamma $'s denote the
polarization of the final $\rho ^{-}$ (and in the SM of the
intermediate off-shell
$W^-$ boson), either
``L=longitudinal'' or
``T=transverse''; superscripts denote ``$\pm $ for
sum/difference of
the $\nu _{L\ }$versus $\nu _R$ contributions''.
The remaining partial-width parameters are defined by
\begin{equation}
\zeta \equiv (\Gamma _L^{-}-\Gamma _T^{-})/(
{\cal S}_\rho \Gamma ), \hspace{2pc} \sigma \equiv (\Gamma
_L^{+}-\Gamma
_T^{+})/(
{\cal S}_\rho \Gamma ).
\end{equation}
The definiton for $\sigma$ in Eq.(4) implies that
\newline $ \tilde{\sigma} \equiv$ (Prob $\rho $ is $\rho _L$)
$
-
$ (Prob $\rho $ is $\rho _T$), where
$$
\tilde{\sigma} = {\cal S}_\rho \sigma ,
$$
is the analogue of the neutrino's chirality parameter in
Eq.(3).
Thus, the parameter $\sigma$, or $\tilde{\sigma}$, measures
the
degree of polarization of the emitted rho.  If the exchange
is, indeed, via an off-shell $W$-boson, $\sigma$ measures the
polarization of the $W$-boson.

To describe the interference between the ${\rho / W }_L$ and
${\rho / W }_R$ amplitudes,
we define
\begin{equation}
\begin{array}{c}
\omega \equiv I_{
{\cal R}}^{-}\ /({\cal R}_\rho \Gamma ), \hspace{2pc}  \eta
\equiv I_{
{\cal R}}^{+}\ /({\cal R}_\rho \Gamma ) \\ \omega ^{\prime
}\equiv I_{
{\cal I}}^{-}\ /({\cal R}_\rho \Gamma ), \hspace{2pc} \eta
^{\prime }\equiv
I_{{\cal I}%
}^{+}\ /({\cal R}_\rho \Gamma )
\end{array}
\end{equation}
where the measureable $LT$-interference intensities are
\begin{equation}
\begin{array}{c}
I_{{\cal R}}^{\pm }=\left| A(0,-\frac 12)\right| \left| A(-
1,-\frac
12)\right| \cos \beta _a \pm \left| A(0,\frac 12)\right|
\left| A(1,\frac
12)\right| \cos \beta _a^R  \\
I_{{\cal I}}^{\pm }=\left| A(0,-\frac 12)\right| \left| A(-
1,-\frac
12)\right| \sin \beta _a \pm \left| A(0,\frac 12)\right|
\left| A(1,\frac
12)\right| \sin \beta _a^R
\end{array}
\end{equation}
Here $\beta _a\equiv \phi _{-1}^a-\phi _0^a$, and $\beta
_a^R\equiv \phi
_1^a-\phi _0^{aR}$\ are the measurable phase differences of
of the
associated helicity amplitudes
$A(\lambda_{\rho},\lambda_{\nu})=\left|
A\right| \exp \iota \phi $.

In the standard model with only a $(V-A)$ coupling and
$m_{\nu} = 0$, these parameters all equal one except for the
two parameters directly sensitive to leptonic $T$ violation
which
vanish,i.e. ${\omega ^{\prime } } = 0 $ and ${\eta ^{\prime }
} = 0$.

The hadronic factors ${\cal S}_\rho $ and ${\cal R}_\rho
$ have
been explicitly inserted into the definitions of the semi-
leptonic decay
parameters, so that quantities such as ${q_\rho }^2={m_\rho
}^2$ can be
smeared over in application due to the finite $\rho $ width.
For the $\rho $
mode they are given by
\begin{equation}
{\cal S}_\rho =\frac{1-2\frac{m_\rho
^2}{m^2}}{1+2\frac{m_\rho ^2}{m^2}} ,
\hspace{2pc}
{\cal R}_\rho =\frac{\sqrt{2}\frac{m_\rho }m}{1+2\frac{m_\rho
^2}{m^2}} .
\end{equation}
These factors numerically are $({\cal
S},{\cal R})_{\rho
,a_1,K^{*}}=0.454,0.445;-0.015,0.500;0.330,0.472$.

{\bf Sensitivity of semi-leptonic parameters}

The numerical values of ``$\xi, \zeta, \sigma, \ldots $"
are very
distinct for different unique Lorentz couplings, see Table 1
and the tables in Ref.[1].

\begin{table*}[htb]
\setlength{\tabcolsep}{2.pc}
\newlength{\digitwidth} \settowidth{\digitwidth}{\rm 0}
\catcode`?=\active \def?{\kern\digitwidth}
\caption{Values of the measureable polarized-partial-widths
$\Gamma$
for $\rho_{L,T} \nu_{L,R}$ final states for unique Lorentz
couplings:}
\label{tabpre3}
\begin{tabular*}{\textwidth}{@{}lllll}
\hline
\cline{2-3} \cline{4-5}
                 & \multicolumn{1}{l}{$V \mp A$}
                 & \multicolumn{1}{l}{$S \pm P$}
                 & \multicolumn{1}{l}{$f_M + f_E$}
                 & \multicolumn{1}{l}{$f_M - f_E$}         \\
\hline
{$\bf  \; \; Analytic$}                    &
&

&

&

\\
$\Gamma _L^{-} / \Gamma$
& $\pm \frac{1}{2} (1 + {\cal S}_\rho ) $
&
$\pm 1
$            &  $\frac{ \rho^2 }{2 \tau^2 + \rho^2}$
&
$ - \frac{1}{3}
$
\\
$\Gamma _T^{-} / \Gamma$                           & $\pm
\frac{1}{2} (1 - {\cal S}_\rho )  $
&
$ 0
$           & $ \frac{ 2 \tau^2 }{2 \tau^2 + \rho^2}$
&
 $ - \frac{2}{3}                                 $
\\
$\Gamma _L^{+} / \Gamma$                          & $
\frac{1}{2} (1 + {\cal S}_\rho )$
&
$ 1
$            &  $\frac{ \rho^2 }{2 \tau^2 + \rho^2}$
&
$    + \frac{1}{3}                                 $
\\
$\Gamma _T^{+} / \Gamma$                          & $
\frac{1}{2} (1 - {\cal S}_\rho )$
&
$ 0
$            &  $\frac{ 2 \tau^2 }{2 \tau^2 + \rho^2}$
&
$    + \frac{2}{3}                                 $
\\
{$\bf  \; \; Numerical$}       &
&

&

&

\\
$\Gamma _L^{-} / \Gamma$                              & $\pm
0.7(\pm 0.5)$ & $
\pm 1  $

& $                  0.0(0.2)
$     &
 $ -0.3                             $
\\
$\Gamma _T^{-} / \Gamma$                              & $\pm
0.3(\pm 0.5)$      &  $
0   $

&  $1.0(0.8)
$
&
$-0.7$
\\
$\Gamma _L^{+} / \Gamma$                               & $
0.7(0.5)$
&  $ 1
$
& $ 0.0(0.8)
$         &
$ +0.3 $
\\
$\Gamma _T^{+} / \Gamma$                                & $
0.3(0.5)$
&  $ 0
$
&  $ 1.0(0.8)
$         &
$ +0.7 $
\\
\hline
\multicolumn{5}{@{}p{120mm}}{}
\end{tabular*}
\end{table*}

In contrast to the purely leptonic modes, the tau
semi-leptonic modes are qualitatively distinct since they
enable a second-stage spin-correlation.  From existing
results, a quantitative comparison
with the ideal sensitivity
in the purely leptonic case is possible if we assume an
arbitrary mixture of
$V$ and $A$ couplings with $m_\nu =0.$ Then the semi-leptonic
chirality parameter $%
\xi _\rho $ and the chiral polarization parameter $\xi
_{Lepton}$ can be
compared since then they both equal $(|g_L|^2-
|g_R|^2)/(|g_L|^2+|g_R|^2)$.
By using $I_4$ to obtain $\xi _\rho $ from $\{\rho ^{-}\rho
^{+}\}$, the
statistical error [2] is $\delta (\xi _\rho )=0.006$ at
$M_Z$.
This is a
factor of $8$ better than the pure leptonic mode's $\delta
(\xi
_{Lepton})=0.05$ error [6] from averaging over the $\mu $
and
$e$ modes and
using $I_3(E_1,E_2,\cos \psi _{12})$ where $\psi_{12}$ is the
openning angle between the two final charged leptons in the
cm-
frame. A complete determination
of the purely
leptonic parameters for $\tau ^{-}\rightarrow \mu ^{-}
\bar{\nu
_\mu} \nu _\tau $
will require a difficult measurement of the $\mu $
polarization,
see
Fetscher[7].

\section{TESTS FOR ``NEW PHYSICS" IN TAU DECAYS}

``New physics'' due to additional Lorentz
couplings in ${J^{Charged}}_{Lepton}$ can show up
experimentally because of
its interference with the $(V-A)$ part which, we assume,
arises as predicted
by the standard lepton model. Therefore, c.f. Eqs.(5),
imaginary parts can be directly measureable.

{\bf Tests for leptonic non-CKM $CP/T$ violation:}

(i) Barred parameters $ \bar{\xi}, \bar{\zeta}, \ldots $ have
[1] the analogous
definitions for the CP conjugate modes, $\tau
^{+}\rightarrow
\rho
^{+}\bar{\nu}, \ldots $. Therefore, any $
\bar{\xi} \neq \xi,
\bar{\zeta} \neq \zeta, \ldots $ $ \Longrightarrow $ CP is
violated: As was shown in [8], if
only $\nu_L$ and $\bar\nu_R$ exist,  there are two simple
tests for ``non-CKM-type" leptonic CP violation in $\tau
\rightarrow \rho \nu$ decay.  Normally a CKM leptonic-phase
will contribute equally at tree level to both the $\tau^-$
decay amplitudes (for exceptions see footnotes
14, 15 in [8]).  These two tests follow because
by CP invariance $B\left(
\lambda_{\bar\rho},\lambda_{\bar\nu}\right)  = \gamma_{CP}
A\left(-
\lambda_{\bar\rho},-\lambda_{\bar\nu}\right) $.  So the two
tests for leptonic CP
violation are: %
\beq
\beta _a=\beta _b \hspace{2pc} {\bf first \hspace*{.4pc}
test}
\eeq
where $\beta _a=\phi _{-1}^a-\phi _0^a$, $\beta _b=\phi _1^b-
\phi
_0^b$, and
\beq
r_a=r_b \hspace{2pc} {\bf second  \hspace*{.4pc} test}
\eeq
where%
\beq
r_a=\frac{|A\left( -1,-\frac 12\right) |}{|A\left( 0,-\frac
12\right)
|},r_b=%
\frac{|B\left( 1,\frac 12\right) |}{|B\left( 0,\frac
12\right) |}
\eeq
Sensitivity levels for $\tau \rightarrow \rho \nu$ and
$\tau \rightarrow  a_1 \nu$ decays are to about $0.05$ to $
0.1\%$ for $r_a = r_b$, and to about $1^o$ to $3^o$ for
$\beta_a = \beta_b$ at $10 GeV$ and at $4 GeV$ without using
polarized $e^- e^+$ beams,
see [6,1].

(ii) Primed parameters $ \omega ^{\prime } \neq 0 $ and/or $
\eta
^{\prime }
\neq 0
\Longrightarrow  \tilde{T}_{FS} $ is violated:  There is a
basic theorem in quantum mechanics that measurement of a
non-real helicity amplitude implies a violation of $
\tilde{T}_{FS} $ invariance when a first-order perturbationin
an ``effective" hermitian Hamiltonian is reliable.  So a
violation of $ \tilde{T}_{FS} $ invariance would imply either
(i) a significant final state interaction between the final
(L, versus T polarized) $\rho$ and the $\nu_{\tau}$, (ii) a
violation of canonical $T$ invariance, or (iii) both (i) and
(ii).  In quantum field theory, nothing forbids either (i)
and/or
(ii) from occurring in $\tau$ decays, so it remains something
to be tested by on-going and future experiments. {\bf Note:}
Canonical CPT invariance implies only equal {\bf total
widths} between a
particle and its antiparticle.  Canonical CPT invariance does
not imply equal partial widths between CP-conjugate decay
modes of a particle and its antiparticle.  Indeed, in nature
in the
kaon system the {\bf partial widths} of the neutral kaons do
differ
for
the particle and the antiparticle.

Note that the trigonometric structure of Eqs.(6)
implies the two constraints
\begin{equation}
(\tilde{\eta} \pm \tilde{\omega} )^2+(\tilde{\eta^{\prime }}
\pm
\tilde{\omega^{\prime }} )^2=\frac
14[(1 \pm \xi )^2-( \tilde{\sigma} \pm \tilde{\zeta} )^2]
\end{equation}
or

$$
2|\eta ^{^{\prime }}\pm \omega ^{^{\prime }}|=\sqrt{(1 \pm
\xi
)^2-
(\tilde{\sigma} \pm
\tilde{\zeta} )^2-4(\tilde{\eta} \pm \tilde{\omega} )^2}
$$
among the $\eta ,\eta ^{^{\prime }},\omega ,\omega ^{^{\prime
}}$ parameters which test for leptonic $ \tilde{T}_{FS} $
violation.  Consistency, i.e. unitarity, requires the
argument
of the square root must be non-negative. To test for leptonic
$ \tilde{T}_{FS} $ violation, besides the $\omega$
parameter which can be measured from $I_4$ in both the $\rho$
and $a_1$
modes, there is the $\eta ^{\prime } $ parameter which can be
obtained
from $I_5$ in both the $\rho$ and $a_1$ modes.  Also there
are
the $\eta$ and $\omega ^{\prime } $ parameters which only
appear in
S2SC
distributions for the $a_1$ modes.

For $10^7$ ($\tau^- , \tau^+ $) pairs at 10 GeV:
from the $ \lbrace \rho^- , \rho^+ \rbrace $ mode and using
the
four-variable distribution $I_4$, the ideal statistical
percentage errors are for $ \omega$, $ 0.6 \%
$. From the $ \lbrace
a_{1}^- , a_{1}^+ \rbrace $ mode: using $I_5^{-}$ the errors
are
for
$ \eta $, $0.6 \% $; using $I_7 $ for $ \eta^{'} $, $ 0.013
$; and using $ I_7^{-} $ for $ \omega ^{\prime } $, $ 0.002
$.
Therefore[1], since these results are more sensitive but use
more angular variables than those included in the simple
spin-correlation function $I_4$, there are better observables
for searching for leptonic T violation than the simple $I_4$
distribution considered in Ref.[2].

{\bf Tests for violation of $(V-A)$ symmetry:}

The most general Lorentz coupling for \hskip
1em  $\tau^{-
}\rightarrow \rho
^{-}\nu _{L,R}$ is
\beq
\rho _\mu ^{*}\bar u_{\nu _\tau }\left( p\right) \Gamma ^\mu
u_\tau
\left(
k\right)
\eeq
where $k_\tau =q_\rho +p_\nu $. It is convenient to treat the
vector
and
axial vector matrix elements separately. In Eq.(12)
$$
\Gamma _V^\mu =g_V\gamma ^\mu +
\frac{f_M}{2\Lambda }\iota \sigma ^{\mu \nu }(k-p)_\nu   +
\frac{g_{S^{-}}}{2\Lambda }(k-p)^\mu +\frac{g_S}{2\Lambda
}(k+p)^\mu
+%
\frac{g_{T^{+}}}{2\Lambda }\iota \sigma ^{\mu \nu }(k+p)_\nu
$$
\beq
\Gamma _A^\mu =g_A\gamma ^\mu \gamma _5+
\frac{f_E}{2\Lambda }\iota \sigma ^{\mu \nu }(k-p)_\nu \gamma
_5
+
\frac{g_{P^{-}}}{2\Lambda }(k-p)^\mu \gamma
_5+\frac{g_P}{2\Lambda }%
(k+p)^\mu \gamma _5  +\frac{g_{T_5^{+}}}{2\Lambda }\iota
\sigma ^{\mu \nu
}(k+p)_\nu \gamma _5
\eeq

The parameter
$%
\Lambda =$ ``the effective-mass scale of new physics''. In
effective field
theory
this
is the scale at which new particle thresholds are expected to
occur or where the theory becomes non-perturbatively
strongly-interacting so as to overcome perturbative
inconsistencies.  It can also be interpreted as a measure of
a
new compositeness scale.  In old-fashioned renormalization
theory
$\Lambda$  is the scale at
which the calculational methods and/or the principles of
``renormalization''
breakdown. While some terms of the
above form do occur as higher-order perturbative-corrections
in
the standard model,
such SM
contributions  are ``small'' versus the sensitivities of
present tests in
$\tau$ physics
in the analogous cases of the $\tau$'s neutral-current and
electromagnetic-current
couplings, c.f. [3].  For charged-current couplings, the
situation
should be the
same.

Without additional theoretical or
experimental
inputs, it is not possible to select what is the "best"
minimal set of couplings for
analyzing the structure of the tau's charged current.  For
instance, by  Lorentz
invariance, for the $\rho$, $a_1$, $K^*$ modes there are the
equivalence theorems that for the
vector
current%
\ber
S\approx V+f_M, & T^{+}\approx -V+S^{-}
\eer
\noindent
and for the axial-vector current
\ber
P\approx -A+f_E, & T_5^{+}\approx A+P^{-}
\eer
There are similar but different equivalences for the $\pi$,
$K$ modes [2]. Therefore, from the perspectives of ``clear
thinking" and of searching for the fundamental dynamics, it
is important to investigate what limits can be set for a
variety of Lorentz structures (including $S^{\pm}$,
$P^{\pm}$, $T^{\pm}$, and ${T_5}^{\pm}$) and not just for a
kinematically minimal, but theoretically prejudiced, set.

Table 2 gives the limits on $\Lambda$ in $GeV$ for real
$g_i$'s from the $\rho$ and $a_1$ modes.  Note that effective
mass scales of $\Lambda \sim 1-2 TeV$ can be probed at $10
GeV$, and at $4 GeV$ for the $(S+P)$ and the $f_M + f_E$
couplings. For determination of ideal statistical errors, we
assume  $10^7$ ($\tau ^{-}\tau ^{+}$%
) pairs at $10GeV$ and separately at $4GeV$; at $M_Z$ we
assume $10^7$
$Z^o$'s with BR($Z^o\rightarrow \tau ^{-}\tau
^{+}$)$=0.03355$; $BR_{\rho}=24.6 \%$, $BR_{a_1}=18\%$ for
the sum of neutral/charged $a_1$ modes, and $BR_{\pi}=11.9
\%$.

We list the ideal
statistical error for the presence of an additional $V+A$
coupling as an error $\delta (\xi _A)$ on the chirality
parameter $\xi _A$
for $\tau ^{-}\rightarrow A^{-}\nu $. Equivalently, if one
ignores possible
different L and R leptonic CKM factors, the effective lower
bound on an
additional $W_R^{\pm }$ boson (which couples only to right-
handed
currents)
is
$$
M_R=\{\delta (\xi _A)/2\}^{-1/4}M_L
$$
For the $\{\rho ^{-},\rho ^{+}\}$($\{a_1{}^{-},a_1{}^{+}\}$)
mode, from $\delta (\xi _\rho
)=0.0012(0.0018)$ this gives equivalently
$M_R>514GeV(464GeV)$. Probably, $10^8$($\tau ^{-
}\tau ^{+}$)
pairs will be accumulated by a $\tau $/charm factory at
$4GeV$, so all
the potential $4GeV$ bounds might be improved by a factor of
$3.2$.

Table 3 gives the
limits from $\tau \rightarrow \pi \nu$.  Note that the $\pi$
mode is important for separating the $(V-A)$ coupling and the
$(T^+ + {T_5}^+ )$ coupling which cannot be distinguished
from the $\rho$ and $a_1$ modes.  Unfortunately, the present
and potential experimental bounds on $(S^- \pm P^-)$
couplings are exceptionally poor or non-existent from
measurements of the $\pi$, $\rho$ and $a_1$ modes.

\begin{table*}[hbt]
\setlength{\tabcolsep}{1.5pc}
\caption{Limits on $\Lambda$'s from $\tau \rightarrow \rho
\nu,
a_1 \nu$ for
Real Coupling
Constants}
\label{tab:creal1}
\begin{tabular*}{\textwidth}{@{}l@{\extracolsep{\fill}}rrrr}
\hline
                 & \multicolumn{2}{l}{$\lbrace \rho^{-},
\rho^{+}
\rbrace$ mode}
                 & \multicolumn{2}{l}{$\lbrace a_{1}^{-},
a_{1}^{+}
\rbrace$ mode}
\\
\cline{2-3} \cline{4-5}
                 & \multicolumn{1}{r}{At $M_Z$}
                 & \multicolumn{1}{r}{10, or 4 GeV}
                 & \multicolumn{1}{r}{At $M_Z$}
                 & \multicolumn{1}{r}{10, or 4 GeV}
\\
\hline
$V+A$, for $\xi_A$                    & $0.006$         &
$0.0012$ &
$0.010$ &
$0.0018$ \\
$S+P$, for $\Lambda$                & $310 GeV$   & $1,700$
&  $
64$     & $
350$ \\
$S-P$, for $(\Lambda)^2$          &$(11GeV)^2$& $(25)^2$ &
$(4)^2$  &
$(7)^2,(10)^2$ \\
$f_M+f_E$, for $\Lambda$      & $210 GeV$   & $1,200$   &
$280$   & $1,500$
\\
$f_M-f_E$, for $(\Lambda)^2$&$(9GeV)^2$  & $(20)^2$ &
$(10)^2$& $(24)^2$
\\
\hline
\multicolumn{5}{@{}p{120mm}}{ For the $\rho$ and $a_1$
modes, the
$T^{+}+T_5^{+}$ coupling is equivalent to the $V-A$ coupling;
and $T^{+}-
T_5^{+}$ is equivalent to $V+A$.}
\end{tabular*}
\end{table*}

\begin{table*}[hbt]
\setlength{\tabcolsep}{1.5pc}
\caption{Limits on $\Lambda$'s from $\tau \rightarrow \pi
\nu$:}
\label{tab7}
\begin{tabular*}{\textwidth}{@{}l@{\extracolsep{\fill}}rrr}
\hline
                 & \multicolumn{1}{c}{From $\xi_\pi$:}
                 & \multicolumn{2}{l}{From $\Gamma (\tau
\rightarrow \pi \nu )$ }
 \\
\cline{2-4}
                 & \multicolumn{1}{c}{ $\vert g_i /  g_L
\vert^2 $}
                 & \multicolumn{1}{r}{$\vert g_i /  g_L
\vert^2 $}
                 & \multicolumn{1}{r}{$2 Re( {g_L}^* g_i )$}
\\
\hline
$V+A$, for $\xi_\pi$                                   &
$0.015,0.004,0.009$
&
$0.014$             & $            $ \\
$S+P, T^+ + {T_5}^+$, for$\Lambda$     &$           $
&
$     $                  & $127GeV$ \\
$S-P, T^+ -
{T_5}^+$,for$(\Lambda)^2$&$(10GeV)^2,(21GeV)^2,(13GeV)^2$&
$(<1GeV)^2$   & $   $ \\
$S^{-} +P^{-}$, for$\Lambda$                 &$           $
&
$     $                  & $<1GeV$ \\
$S^{-} -P^{-}$,for$(\Lambda)^2$           &$(<
1GeV)^2,(1.6GeV)^2,(1GeV)^2$&
$(<1GeV)^2$   & $   $ \\

\\
\hline
\multicolumn{4}{@{}p{100mm}}{}
\end{tabular*}
\end{table*}

\newpage

{\bf Tests for $\tau$ compositeness:}

In analogy with the Pauli anomalous magnetic moment, an
obvious signature for lepton compositeness would be an
additional tensorial coupling. In this regard, it is useful
to first test for the presence of only $\nu_L$ couplings
which would exclude a significant contribution from the $g_{-
} = f_M - f_E$ tensorial coupling. For example, for the $a_1$
and $\rho$ modes there are 3
logically independent tests for only $\nu_L$ couplings: the
chirality parameter $\xi
= 1$, $\zeta = \sigma$, and $\omega = \eta$.  In addition, if
$\tilde T_{FS}$ violation
occurred then the non-zero parameters $\omega^{'} = \eta^{'}$
if there are only $\nu_L$ couplings.

On the other hand, just as in the case of a pure $(V-A)$
coupling, an additional tensorial $g_{+}=f_M+f_E$ coupling
would preserve these 3 signatures for only $\nu _L$
couplings.  But, such a tensorial $g_{+}=f_M+f_E$ coupling
would give non-($V-A$)-values: $\zeta = \sigma \neq 1$ and
$\omega = \eta \neq 1$.  Second, there is
the prediction that for $\Lambda $ large
\begin{equation}
(\zeta -1)=(1-\omega )\frac gl
\end{equation}
where  the ratio `` $g/l$ '' is a known function [1] of
$m_\rho $ and $%
m_\tau $. Numerically  $(g/l)_{\rho }=0.079$.

These $\nu_L$ signatures and Eq.(16) also
occur for an additional $(S+P)$ coupling but with
the ratio $(g/l)$ replaced by $(a/d)$, which varies from
$5.07$ to $12.1$ across
$(m_{\rho} \pm \Gamma / 2 $.  Fortunately, here the $\pi$
mode can again be used to limit the presence of an additional
$(S+P)$ coupling versus an $g_+ = f_M + f_E$ coupling, see
second-line in Table 3.

\section{CONCLUSIONS}

In the future, there will be a major theoretical and
experimental effort to determine whether or
not the Higgs-mechanism is due to a fundamental scalar
elementary field or, instead, due to a macroscopic $f
\bar{f}$ mechanism a'la the BCS theory in superconductivity.
Of equal importance are the deep questions
raised by the existence of the two massive charged-leptons,
the $\mu$
and the $\tau$.  In the case of the leptons, it is also
important to study whether they are truly elementary or only
macroscopic.  Besides searching for rare/forbidden decays, a
direct way in which to proceed is to search for violations of
$CP$, $T$, and $(V-A)$ symmetries in $\tau$ decays.
\newline
\newline

{\bf Acknowledgments:}
This work was partially supported by U.S. Dept. of Energy
Contract No. DE-FG 02-96ER40291.  We thank the organizers and
participants for an exciting and helpful WHEPP4 workshop in
Calcutta.


\begin{thebibliography}{99}
\bibitem{1} C.A. Nelson, Phys. Rev. {\bf D53}, 5001(1996).
Related works on polarization and spin-correlation tests in
$\tau$ physics include Y.-S. Tsai, Phys. Rev. {\bf D4},
2821(1971);
S.Y. Pi and A.I. Sanda, Ann. Phys. (N.Y.) {\bf 106},
171(1977); J. Babson and E. Ma, Phys. Rev. {\bf D26}, 2497
(1982);  J.H. Kuhn and F. Wagner, Nuc. Phys. {\bf B236},
16(1984);
A. Rouge, Z. Phys {\bf C48},  75(1990);  K. Hagiwara,
A.D. Martin, and D. Zeppenfeld, Phys. Lett. {\bf  B235},
198(1990); R. Alemany, N. Rius, J. Bernabeu, J.J. Gomez-
Caenas, A. Pich, Nuc. Phys. {\bf B379}, 3(1992);
M. Davier, L. Duflot,
F.Le
Diberder, and A. Rouge, Phys. Lett. {\bf B306},  411(1993);
H. Thurn and H. Kolanski, Z. Phys. {\bf C60}, 277(1993);
B.K.
Bullock, K. Hagiwara, and A.D. Martin, Nuc. Phys. {\bf B359},
499 (1993); A. Pich and J. P. Silva, Phys. Rev. {\bf D52},
4006(1995); J.H. Kuhn, E. Mirkes, and M. Finkemeier,
hep-ph/9511268.
\bibitem{2} C.A. Nelson, Phys. Lett. {\bf B355}, 561(1995).
\bibitem{3}  S.M. Barr and W. Marciano, in {\em CP
Violation},
C. Jarlskog(ed), World Sci., Singapore, 1989; W. Bernreuther,
U.
Low, J.P. Ma and O. Nachtmann, Z. Phys. {\bf C43},
117(1989); C.A. Nelson, Phys. Rev. {\bf D43}, 1465(1991); S.
Goozovat and C.A. Nelson, Phys. Letters {\bf B267},
128(1991), Phys. Rev. {\bf D44}, 2818(1991); J. Bernabeu, N.
Rius, and A. Pich, Phys. Lett.
{\bf B257},  219(1991);  J.A. Grifols and A. Mendez,
Phys. Lett. {\bf B255},  611(1991); and  R. Escribano and E.
Masso, Phys. Lett. {\bf B301}, 419(1993); UAB-FT-317; W.
Bernreuther and O. Nachtmann, hep-ph/960331; U. Mathanta,
hep-ph/9604380; W. Bernreuther, A. Brandenburg and P.
Overmann, hep-ph/9608364.
\bibitem{3a} C.A. Nelson, Phys. Rev. {\bf D41}, 2805(1990);
p. 259 {\em Results and Perspectives in Particle Physics},
M. Greco(ed), Editions Frontiers, France 1994; C. A. Nelson,
B. Kress, M. Lopes and T. McCaulley, paper in preparation.
\bibitem{4} C.A. Nelson, hep-ph/9608439, and a paper in
preparation.
\bibitem{5} C.A. Nelson, Phys. Rev. {\bf D40}, 123(1989);
{\bf D41}, 2327(1990)(E).
\bibitem{6} W. Fetscher, Phys. Rev. {\bf D42},  1544(1990).
\bibitem{7} C.A. Nelson, H.S. Friedman, S. Goozovat, J.A.
Klein,
L.R. Kneller, W.J. Perry, and S.A. Ustin, Phys. Rev. {\bf
D50},
4544(1994); C.A. Nelson, SUNY BING 7/19/92; p.353 {\em Proc.
of
the Second Workshop on Tau Lepton Physics}, K.K. Gan (ed),
World Sci., Singapore, 1993. Recent work on CP
violation in tau decays includes U. Kilian, J.G. Korner, K.
Schilcher and Y.L. Wu, Z. Phys. {\bf C62}, 413(1994); S.Y.
Choi, K. Hagiwara and
M. Tanabashi, Phys. Rev. {\bf D52}, 1614(1995); Y.-S. Tsai,
Phys. Rev. {\bf D51}, 3172(1995); T. Huang, W. Lu and Z. Tao,
hep-ph/9609220.

\end{thebibliography}
\end{document}